\documentclass[reviewcopy]{elsarticle}
\usepackage[reviewcopy]{adndt}
\usepackage{longtable}
 \usepackage{array}
 \usepackage{multirow}
 \setlongtables
\usepackage{hyperref}
\usepackage{pdflscape} 
\usepackage{fancyheadings}
\usepackage{subfig}
\usepackage{amsmath}
\usepackage{footnote}
\usepackage{rotating}
\usepackage{lineno}
\usepackage{appendix}

\usepackage{amsmath}

\usepackage{amssymb}

\biboptions{square,sort&compress}
\bibpunct[]{[}{]}{,}{n}{}{;}
\begin{document}

\title{
Tables of Neutron Thermal Cross Sections, Westcott Factors, Resonance Integrals, Maxwellian Averaged Cross Sections, Astrophysical Reaction Rates, and $r$-process Abundances Calculated from the ENDF/B-VIII.1, JEFF-3.3, JENDL-5.0, BROND-3.1, and CENDL-3.2  Evaluated Data Libraries
}
\cortext[cor1]{Corresponding author}
\author[label1]{B. Pritychenko\corref{cor1}}
\ead{pritychenko@bnl.gov}
\address[label1]{National Nuclear Data Center, Brookhaven National Laboratory, \\ Upton, NY 11973-5000, USA}

\date{\today}

\begin{abstract}
We present calculations of  neutron thermal cross sections, Westcott factors, resonance integrals, 
Maxwellian-averaged cross sections, astrophysical reaction rates, and solar system $r$-process abundances using the latest data from the 
major evaluated nuclear libraries for 849 ENDF target materials. 
The recent release of ENDF/B-VIII.1 library, progress in $^{252}$Cf(SF) evaluation,  extensive analysis of newly-evaluated neutron reaction cross sections, neutron covariances, 
and improvements in data processing techniques motivated us to calculate the nuclear industry and 
neutron physics parameters, produce $s$-process Maxwellian-averaged cross sections and 
astrophysical reaction rates, extract $r$-process abundances, systematically calculate uncertainties, 
and provide additional insights on currently available neutron-induced reaction data. 
 \end{abstract}


\maketitle


\tableofcontents{}

\section{Introduction}
\label{sec:introduction}
Neutron physics integral values are essential in developing atomic energy, national security, and nuclear astrophysics applications~\cite{11Chad,06Mugh,18Mugh,05Nak,10Pri,12Pri,21Pri}. These values could be extracted from the international collection of Evaluated Nuclear Data File (ENDF) reaction libraries: ENDF/B-VIII.1 (USA), JEFF-3.3 (European Union), JENDL-5.0 (Japan), CENDL-3.2 (China), and BROND-3.1 (the Russian Federation)~\cite{25Nob,20Plo,23Iwa,16Blo,20Ge}. The evaluated libraries represent the best efforts of national nuclear data programs and can be described as ``national" evaluated nuclear data libraries.  To provide users access to the latest neutron physics data, we have processed the data files, 
calculated the integral values, and analyzed the results. 

The national libraries contain complete data sets for 849 neutron target materials (nuclei), and 254 target materials are present in the neutron sublibraries of the five libraries. The shown below list of common 254 neutron targets illustrates comprehensive nuclear reaction evaluation efforts around the globe and explains the interest in the evaluated data sets for multiple applications: $^{1-3}$H, $^{3,4}$He, $^{6,7}$Li, $^{9}$Be, $^{10,11}$B, C/$^{12,13}$C, $^{14}$N, $^{16}$O, $^{19}$F, $^{23}$Na, $^{24-26}$Mg, $^{27}$Al, $^{28-30}$Si, $^{31}$P, $^{32-34,36}$S, Cl/$^{35,37}$Cl, K/$^{39-41}$K, Ca/$^{40}$Ca, $^{46-50}$Ti, V/$^{50,51}$V, $^{50,52-54}$Cr, $^{55}$Mn, $^{54,56-58}$Fe, $^{59}$Co, $^{58,60-62,64}$Ni, $^{63,65}$Cu, $^{64,66-68,70}$Zn, $^{69,71}$Ga, $^{70,72-74,76}$Ge, $^{75}$As, $^{76-80,82}$Se, $^{83-86}$Kr, $^{85-87}$Rb, $^{88-90}$Sr, $^{89,91}$Y, $^{90-96}$Zr, $^{93,95}$Nb, $^{92,94-100}$Mo, $^{99}$Tc, $^{99-105}$Ru, $^{103,105}$Rh, $^{105,108}$Pd, $^{107,109}$Ag, Cd/$^{113}$Cd, $^{113,115}$In, $^{112,114-120,122,124,126}$Sn, $^{121,123-125}$Sb, $^{130}$Te, $^{127,129-131,135}$I, $^{129,131-136}$Xe, $^{133-135,137}$Cs, $^{130,132,134-138}$Ba,  $^{139}$La,  $^{136,138,140-142,144}$Ce,  $^{141}$Pr, $^{142-148,150}$Nd, $^{147-149}$Pm, $^{144,147-152,154}$Sm, $^{151,153-155}$Eu, $^{152,154-158,160}$Gd, $^{164}$Dy, $^{165}$Ho, $^{174,176-180}$Hf, $^{181}$Ta, $^{180,182-184,186}$W, $^{197}$Au, $^{204,206-208}$Pb, $^{209}$Bi, $^{232}$Th, $^{232-238,240}$U, $^{236-239}$Np, $^{236-242,244,246}$Pu, and $^{241-243}$Am. 
The analysis of common target materials shows that the national libraries provide recommended reaction data values for stable and long-lived radioactive nuclei.  Far from the valley of stability, where half-lives are measured on a millisecond scale and no experimental data are available, the TALYS calculated TENDL library~\cite{12Kon} is recommended. The TENDL values have been previously used to deduce Maxwellian-averaged cross sections (MACS), astrophysical reaction rates, $r$-process abundances~\cite{20Pri}, and astrophysical data sets~\cite{25Roc}.

In the present work, we consider  the five major libraries neutron elastic  scattering (n,n), capture (n,$\gamma$), and fission (n,f)  
reactions and several important integral values for nuclear science and technology applications.
Integral values of thermal neutron cross sections ($\sigma^{2200}$), Westcott factors ($g_w$), 
resonance integrals (RI),  Maxwellian-averaged cross 
sections  (MACS or $\sigma^{Maxw}$), astrophysical reaction rates ({\it R(T$_9$)}), and solar system $r$-process abundances were calculated in a
a systematic approach for Z=1-100 nuclei (materials) using the Doppler broadened nuclear 
reaction data files. 

We analyzed the results and produced recommendations consulting with the {\it Atlas of Neutron Resonances} reference book~\cite{06Mugh,18Mugh}, experimental nuclear reaction data (EXFOR) Library~\cite{X4,EXFOR}, $^{252}$Cf(SF) evaluation~\cite{24Neu}, and ASTRAL database~\cite{23Ves} as benchmarks. Furthermore, International Evaluation of Neutron Cross Section Standards~\cite{09Car,15Car,18Car} is an evolving list of evaluated neutron reactions on H, $^{3}$He, $^{6}$Li, $^{10}$B, C, $^{197}$Au, $^{235,238}$U, and $^{239,241}$Pu, and it is often used for pinpointing neutron cross sections. The neutron standards are included in the ENDF/B-VIII.1 library as a sublibrary, and evaluated neutron sublibraries incorporate its data. In addition, the International Reactor Dosimetry and Fusion File, IRDFF-II~\cite{20Tr}, provides recommended dosimetry cross sections for 70 neutron target materials (isotopes and elements) that are used in ENDF/B evaluations.

The current analysis includes 849 isotopic and elemental neutron
evaluations of importance to various applications.  This paper uses 
ENDF neutron cross section covariance data, to
provide uncertainty estimates. An extensive discussion on integral value calculations,
potential implications, and further recommendations are presented. Since the current
calculations were completed by September 2024,  only nuclear data library versions 
available before this date were considered.

\section{Evaluated Nuclear Reaction Data Libraries}
\label{sec:libraries}
The increasing energy demand and clear connection between the usage of fossil fuels and climate change make a strong case for using nuclear power.  Presently, $\sim$20\% of U.S. electrical power is generated by atomic power plants~\cite{10Phy}, and high-quality nuclear data are needed for their operation. The nuclear reaction data files are stored in the databases or libraries using the internationally adopted ENDF-6 format~\cite{11He}.  
This format is maintained by the Cross Section Evaluation Working Group (CSEWG)~\cite{CSEWG}. ENDF is a core nuclear reaction library containing evaluated (recommended) cross sections,  neutron spectra, angular distributions, fission product yields, thermal neutron scattering, photo-atomic, and other data, with
emphasis on neutron-induced reactions. 

The CSEWG and the U.S. Nuclear Data Program (USNDP)~\cite{CSEWG,USNDP,06Pri}  coordinate the evaluation and dissemination of nuclear reaction data in the USA,  and the Working Party on International Nuclear Data Evaluation Co-operation (WPEC)~\cite{WPEC}  and the International Atomic Energy Agency (IAEA)~\cite{04Hum} worldwide. Broad international effort led to the development of a variety of ENDF-6 formatted libraries: ENDF/B in the USA~\cite{06Chad,01Cse}, JEFF in Europe~\cite{11Kon}, JENDL in Japan~\cite{02Shi,11Shi},  CENDL in China~\cite{11Zhi}, ROSFOND and BROND in Russian Federation~\cite{07Zab,BROND} and, USA, European Union, Japan, China, and the Russian Federation continue to invest heavily in evaluated nuclear reaction libraries research and development.

\subsection{Evaluated Neutron Cross Section Data}

Frequently, ENDF cross sections are represented in pointwise (energy-dependent or ``continuous") and GroupWise (averaged over broad energy interval) formats~\cite{11He,10Mac}. 
The first representation is well suited for nuclear physics applications while the second is often used in reactor physics calculations.  In this work, we will consider several uses of evaluated data for slow and fast neutrons 
and compare integral values with the EXFOR database~\cite{X4,EXFOR},  {\it Atlas of Neutron Resonances} reference book~\cite{06Mugh,18Mugh} and other commonly-used standards and benchmarks~\cite{24Neu,23Ves}. 
 ENDF library resonance energy region cross sections are temperature dependent, and not provided by default. To resolve this shortcoming, codes NJOY~\cite{10Mac} and PREPRO~\cite{07Cul}  
are often used for processing (Doppler broadening) neutron cross sections within the ENDF range of energies from 10$^{-5}$ eV to 20 MeV, and beyond if data are available. In the present work, ENDF data were preprocessed with 0.1$\%$ precision within T=293.16$^{\circ}$ K temperature range using code PREPRO~\cite{07Cul}. The latest version of PREPRO 2023 was successful with Doppler broadening of five libraries except for $^{88}$Sr evaluation~\cite{23Pig}  in the ENDF/B-VIII.1 library. The $^{88}$Sr ENDF/B-VIII.1 evaluation was processed with the code NJOY~\cite{10Mac}.

\section{Calculation of Thermal Cross Sections,  Westcott Factors, Resonance Integrals,   Maxwellian-Averages Cross Sections,  Astrophysical Reaction Rates,  and $r$-Process Abundances}
\label{sec:calculation}

We performed calculations of neutron thermal cross sections, Westcott factors, resonance integrals,  
Maxwellian-averaged cross sections and reaction rates 
using the evaluated neutron library data \cite{25Nob,20Plo,23Iwa,16Blo,20Ge}. The definitions of integral values 
 are presented below.

There are multiple descriptions of neutron thermal cross sections in literature: 
\begin{itemize}
\item cross section at 0.0253 eV
\item cross section at 20$^{\circ}$ C or 293.16$^{\circ}$ K
\item cross section for 2200 m/sec neutrons
\end{itemize}
These renditions can be summarized as a cross section of neutrons in thermal equilibrium with the environment. Since neutron cross sections in the lab reference system are tabulated in ENDF evaluations, we Doppler-broaden evaluated reaction data at 293.16 $^{\circ}$ K and extracted these values from the libraries.

Westcott g-factor, $g_w$, is the ratio of thermal  Maxwellian-averaged cross section ($\sigma^{Maxw}$) to the 2200 m/s cross section ($\sigma^{2200}$) \cite{55We,58We}  in the same coordinate system
\begin{equation}
\label{myeq.west1}
g_w = \frac{\sigma^{Maxw}}{\sigma^{2200}}, 
\end{equation}
or
\begin{equation}
\label{myeq.west2}
g_w = \frac{\sigma^{Maxw}}{\sigma^{2200}} = \frac{4}{\upsilon_{0} \sigma_{0} \sqrt{\pi}} \int_{0}^{\infty} \frac{\upsilon^{3}}{\upsilon_{T}^{3}} e^{-(\frac{\upsilon}{\upsilon_{T}})^{2}} \sigma(\upsilon) d\upsilon,
\end{equation}
where $\upsilon_{0}$ (2200 m/sec) is the velocity of a neutron of energy $kT_{0}$ with $T_{0}$ = 293.60$^{\circ}$ K (20.44$^{\circ}$ C), and $\upsilon_{T} = \upsilon_{0} \sqrt \frac{T}{T_{0}}$~\cite{58We}.
The last equation indicates that $g_{w}$-factor is temperature dependent  \cite{55We,58We,inter,99Ho,07Ch} and its value is close to 1 for most nuclei where $\sigma (n,\gamma) \sim \frac{1}{\upsilon}$.

Resonance integrals have been used since the 1950s. They are defined by Lamarsh~\cite{66Lam} in a quasi-physical
interpretation as ``...the resonance integral is equal to the integral of the [effective] absorption [capture]
cross section over the resonance region which is necessary to account for the observed 
the neutron absorption rate in a flux equal to that existing in the absence of the resonances."  The quasi-physical interpretation appears to relate to a physically measurable quantity
(i.e., neutron absorption rate) to some as yet undefined "effective" absorption cross section~\cite{80Bak}. 
The resonance integral is helpful for this reason; physically observable quantities can be
calculated for a given fuel system in terms of its resonance integral. Additionally, the resonance integral represents a single-group representation of ENDF cross sections, and was very useful in early computer calculations.

The epicadmium dilute resonance integral (RI) \cite{06Mugh} for a particular reaction $\sigma_R (E)$ in $1/E$ spectrum is expressed by
\begin{equation}
\label{myeq.res1}
RI =  \int_{E_c}^{\infty} \sigma_R (E) \frac{dE}{E},
\end{equation}
where $E_c$ is determined by cadmium cutoff energy ($E_c$=0.5 eV). The cutoff energy depends on the Cd shield thickness, 
its values are approximately 0.5 and 0.55 eV for $\sim$1 and $\sim$1.5 mm cadmium shields \cite{68De}, respectively. 
 The absorbing nuclide is assumed to be present in such small or diluted quantities that there is no perturbation of the neutron slowing down spectrum  \cite{02Ba}. 
Consequently, the resonance integral is called infinitely dilute RI.

Average cross sections for the Maxwellian spectrum  are defined as 
\begin{equation}
\label{myeq.max1}
\sigma^{Maxw}(kT) =  \frac{\langle \sigma \upsilon \rangle}{\upsilon_T},
\end{equation}
where $\upsilon$ is the relative velocity of neutrons and a target nuclide and $\upsilon_{T}$ is the mean thermal velocity given by 

\begin{equation}
\label{myeq.max2}
\upsilon_{T} = \sqrt{\frac{2kT}{\mu}},
\end{equation}
where $\mu$ is the reduced mass. 

Maxwellian-averaged cross sections (MACS) in the center-of-mass system can be expressed as \cite{10Pri}   

\begin{equation}
\label{myeq.max3}
\sigma^{Maxw}(kT) = \frac{2}{\sqrt{\pi}} \frac{a^{2}}{(kT)^{2}}  \int_{0}^{\infty} \sigma(E^{L}_{n})E^{L}_{n} e^{- \frac{aE^{L}_{n}}{kT}} dE^{L}_{n},
\end{equation}
where $a = m_2/(m_1 + m_2)$, {\it k} and {\it T} are the Boltzmann constant and temperature of the system, respectively and $E$ is an energy of relative motion of 
the neutron concerning the target. $E^{L}_{n}$ is a neutron energy in the laboratory system and $m_{1}$ and $m_{2}$ are masses of 
the neutron and the target nucleus, respectively.

The astrophysical reaction rate, $R$, is defined as $R$ = $N_{A}$$\langle \sigma \upsilon \rangle$, where $N_{A}$ is the Avogadro number. To express reaction rates in [$cm^{3}$/mole s] units, an additional factor of $10^{-24}$ is introduced; $\upsilon_{T}$ is in units of [cm/s] and temperature, $kT$, in units of energy ({\it e.g.} MeV) is related to that in Kelvin ({\it e.g.} 10$^{9}$ K) as $T_{9}$=11.6045$kT$.
\begin{equation}
\label{myeq.rrates}
R(T_{9}) = 10^{-24}N_{A}\sigma^{Maxw}(kT)\upsilon_{T}.
\end{equation}

These equations were used to  calculate integral values using the original evaluated neutron data 
from the ENDF libraries within the typical range of energies.

\subsection{Calculation of ENDF Integral Values}

Originally published computations of Maxwellian-averaged cross sections and astrophysical reaction rates  \cite{10Pri} were based on the Simpson integration method for the linearized ENDF cross sections (MF=3). 
This method provided quality integral values. However, the degree of precision was within $\sim 1 \%$ \cite{05Nak,10Pri}. 
This limitation can be overcome in the linearized ENDF files because cross section value is linearly dependent on energy within a particular bin \cite{10BPri}
\begin{equation}
\label{myeq.int1}
\sigma (E) = \sigma_{1}(E_1) + (E-E_1)\frac{\sigma_2 (E_2) - \sigma_1 (E_1)}{E_2 - E_1}, 
\end{equation} 
where $\sigma (E_1),  E_1$ and  $\sigma (E_2),  E_2$ are pointwise cross sections and energy values for the energy bin.

In the present work, we calculate definite integrals applying the Wolfram Mathematica online integrator \cite{09Math}. 
As an illustration, the computational value of Maxwellian-averaged cross section for a single bin is deduced using Eqs.~\ref{myeq.max3} and ~\ref{myeq.int1}

\begin{eqnarray}
\label{myeq.max4}
\sigma^{Maxw}(kT)&=&\frac{2}{\sqrt{\pi}} \frac{a^{2}}{(kT)^{2}}  \int_{E_1}^{E_2} \sigma(E^{L}_{n})E^{L}_{n} e^{- \frac{aE^{L}_{n}}{kT}} dE^{L}_{n} \\  \nonumber
&=&\frac{2}{\sqrt{\pi}} \frac{a^{2}}{(kT)^{2}} (-\frac{kT}{a^{3} (E_1-E_2)} e^{-\frac{aE^{L}_{n}}{kT}}) \\  \nonumber
&\times& (a^{2}E^{L}_{n}(E_1\sigma_{2} -E_2\sigma_{1} +E^{L}_{n}(\sigma_{1} -\sigma_{2})) + akT(E_1\sigma_{2} -E_2\sigma_{1} + 2E^{L}_{n}(\sigma_{1} - \sigma_{2})) +2(kT)^{2}(\sigma_{1} - \sigma_{2}))  \vert_{E_1}^{E_2} 
\end{eqnarray}

Summing definite integrals for all energy bins and sufficiently dense grids produces an exact integral value for parameters of interest.

\subsection{Calculation of Cross Section Uncertainties}

Major evaluated neutron libraries contain covariance files. In this paper, we will consider the ENDF/B-VIII.1, JEFF-3.3, JENDL-5.0, CENDL-3.2, and BROND-3.1  file 33 (MF33) covariance files~\cite{11He} mostly for structural, fission, and actinide materials, and calculate uncertainties using the error propagation formalism~\cite{92Bev}. The calculated uncertainties supply complementary benchmarks to the templates of expected measurement uncertainties~\cite{New23,Lew23,Van23,Le23} and the Unrecognized Sources of Uncertainties (USU ) in Experimental Nuclear Data IAEA coordinated research project~\cite{Cap20}.

\section{Major Results}
\label{sec:results}
In this section, we present the results and discuss their implications. Neutron thermal cross sections were extracted from the Doppler-broaden ENDF libraries while  Westcott factors, resonance integrals,  
Maxwellian-averaged cross sections and reaction rates for the ENDF/B-VIII.1, JEFF-3.3, 
JENDL-5.0, CENDL-3.2, and BROND-3.1 libraries have been calculated. 
An extensive analysis of the ENDF integral values has been performed in previous publications~\cite{10Pri,12Pri,21Pri}. In this work, we extend our analysis to the latest nuclear data and present results in the appendix tables.

\subsection{Neutron Thermal Cross Sections}
Neutron thermal cross section values for elastic scattering, fission, and capture are shown in Tables~\ref{ThCS2},~\ref{ThCS18},~\ref{ThCS102}. The Doppler-broaden elastic, fission, and capture cross sections are compared with free neutron scattering~\cite{18Mugh}  or scattering~\cite{06Mugh}, fission, and capture cross sections in the recent edition of Atlas of Neutron Resonances reference book.

\subsection{Westcott Factors}
Westcott factor values for elastic scattering, fission, and capture are shown in Tables~\ref{W2},~\ref{W18},~\ref{W102}.  Elastic Westcott factor values are close to 1.129, which is consistent with our previous findings~\cite{12Pri}. The Westcott factors occasionally deviate from unity due to spectra merging issues. Complete calculation of capture and fission Westcott factors reveals that most of them are close to 1 with an exception 
of  non-$1/\upsilon ~\sigma (n,\gamma)$ nuclei: $^{113}$Cd, $^{135}$Xe, $^{149}$Sm,  $^{151}$Eu, $^{176}$Lu, $^{182}$Ta, $^{239}$Pu,  
$^{243}$Am \cite{06Mugh}.  Strong resonances in the thermal energy region, such as 0.29562(30) eV resonance in $^{239}$Pu \cite{06Mugh}, 
are often responsible for Westcott factor temperature variations.   The calculated factors for neutron capture and fission are compared with the Atlas of Neutron Resonances reference book~\cite{18Mugh}. In the Atlas, we notice several typos in the reference book where neutron capture factors for $^{59}$Ni, $^{141}$Ce, $^{152}$Gd, $^{160-161}$Dy, $^{168}$Er, and $^{173}$Lu are not displayed correctly.

\subsection{Resonance Integrals}
Resonance integrals for elastic scattering, fission, and capture are shown in Tables~\ref{RI2},~\ref{RI18},~\ref{RI102}.  The calculated elastic, fission, and capture resonance integrals are compared with elastic, fission, and capture integrals in the Atlas of Neutron Resonances reference book~\cite{18Mugh}. The reference book values calculated by S.~Mughabghab are marked with a {\it C} superscript.

\subsection{Maxwellian-averaged Cross Sections}
Maxwellian-averaged cross sections have been calculated for five temperatures ($kT$) of  8, 25, 30, 90, and 1420 keV. The first four temperatures are commonly used   
in stellar nucleosynthesis modeling, the fifth one closely reproduces $^{252}$Cf neutron fission spectrum. 

The slow-neutron capture ($s$-process) produces $\sim$50 $\%$ of the elements
beyond iron. In this region, neutron capture becomes dominant because of the increasing
Coulomb barrier and decreasing binding energies. This s-process proceeds in the Red Giants
and Asymptotic Giant Branch stars, where neutron temperature ({\it kT}) varies from 8 to 90 keV~\cite{11Kap}.

$^{252}$Cf has been used in nuclear science applications because the californium spectrum shape is close to nuclear reactor neutrons, and it emits a broad variety of fission products. Its fission neutron spectrum data have been compiled in the EXFOR database \cite{EXFOR}. 
Previously we have found that $kT$=1420 keV MACS~\cite{12Pri} deviate from the Mannhart spectrum cross sections by $\sim30\%$~\cite{15Pri}. Further analysis of calculated MACS and experimental data provides an important tool for ENDF libraries' quality assurance.

\subsubsection{{\it kT}=8 keV}
Maxwellian-averaged cross sections for fission and capture are shown in Tables~\ref{Macs18-8},~\ref{Macs102-8}.

\subsubsection{{\it kT}=25 keV}
Maxwellian-averaged cross sections for fission and capture are shown in Tables~\ref{Macs18-25},~\ref{Macs102-25}.  

\subsubsection{{\it kT}=30 keV}
JENDL-5.0 library includes a no-zero theoretical fission cross section above 20 MeV for $^{205}$Pb and no experimental values for lead are available in the EXFOR library.   As a result, MACS  of 1.569E-296 barns is calculated for $kT$=30 keV,  7.641E-105 b for $kT$=90 keV, and 2.331E-14 b at $kT$=1420 keV. A similar situation is also observed in $^{211,222}$Rn, where fission starts from 6.5 and 4 MeV, respectively.

\begin{figure}
\begin{center}
 \includegraphics[width=.65\linewidth]{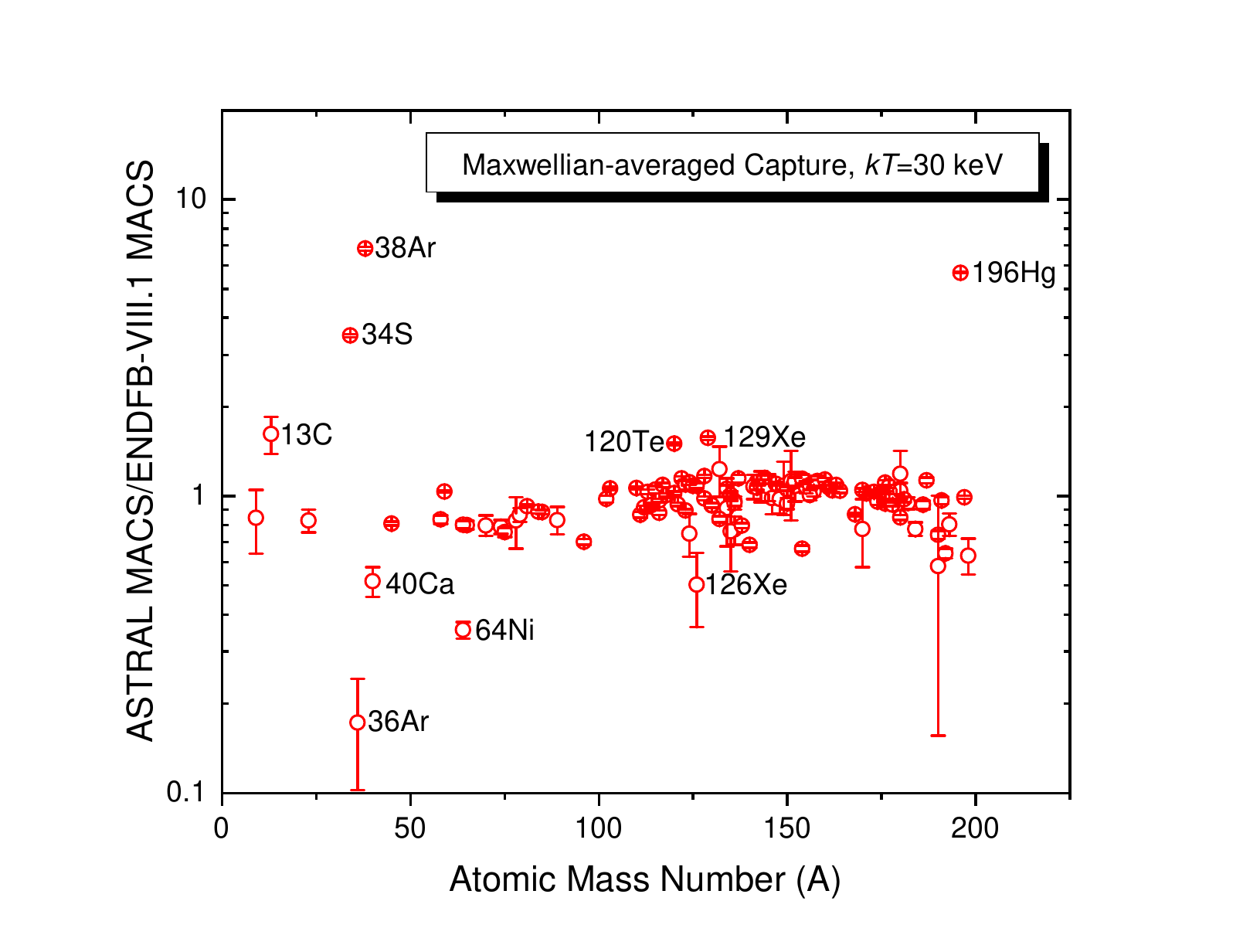}
\end{center}
\caption{ Maxwellian-averaged cross section (MACS) ratio between ASTRAL and ENDF/B-VIII.1 libraries~\cite{23Ves,25Nob}.} \label{fig:Astral}
\end{figure}

Maxwellian-averaged cross section ratios and their numerical values for fission and capture are shown in Fig.~\ref{fig:Astral} and Tables~\ref{Macs18-30},~\ref{Macs102-30}, respectively.  The capture values were compared with the ASTRAL library~\cite{23Ves}, and the detailed analysis is given in the ENDF/B-VIII.1 reference paper~\cite{25Nob}. Further analysis of Table~\ref{Macs102-30} and other tables show the nuclear astrophysics potential of ENDF libraries as a complementary source of evaluated cross sections and reaction rates  \cite{10Pri} for dedicated astrophysics libraries~\cite{23Ves,06Dil}. 

\subsubsection{{\it kT}=90 keV}
Maxwellian-averaged cross sections for fission and capture are shown in Tables~\ref{Macs18-90},~\ref{Macs102-90}.  

\subsubsection{{\it kT}=1420 keV}

Previously Maxwellian-averaged cross sections were calculated using the evaluated data libraries~\cite{12Pri} and Mannhart neutron spectra~\cite{15Pri,06Man}. In the present work, we will provide an update based on the latest nuclear data libraries 
assuming the $^{252}$Cf $\bar{E}$ of 2.13 MeV  and 
employing the formula
\begin{equation}
\label{myeq.252Cf}
\bar{E}_{n} = \frac{3}{2} kT,
\end{equation}
where we deduce  $kT$=1420 keV Maxwellian temperature for an approximation of $^{252}$Cf neutron spectrum \cite{94Tru}. 
The current approximation has limitations and the more advanced formalisms have been proposed \cite{85Gra,98Mad,10Ta}. 
Maxwellian-averaged cross sections for fission and capture are shown in Tables~\ref{Macs18-1420},~\ref{Macs102-1420}.  These results are compared with the EXFOR~\cite{X4,EXFOR} library and Mannhart evaluation~\cite{06Man}.

\subsection{Astrophysical Reaction Rates}

Astrophysical reaction rates were calculated by many authors using the statistical model~\cite{78Wo,00Ra}, TALYS code~\cite{08Go} and evaluated neutron data~\cite{05Nak,10Pri,12Pri,20Pr} approaches. The present work supplements the earlier findings by calculating astrophysical reaction rates and their uncertainties for the whole range of ENDF nuclei at $kT$=8, 25, 30, 90, and 1420 keV.  
An additional calculation for the 1 MK - 10 GK temperature range is available upon request.

\subsubsection{Stellar Enhancement Factors}

The stellar enhancement factors (SEF) are defined as
the ratio of the stellar rate $R^{*}(T_{9})$ relative to the ground state
rate $R^{g.s.}$
\begin{equation}
\label{myeq.SEF}
SEF = \frac{R^{*}(T_{9})}{R^{g.s.}} = \frac{R^{*}(T_{9})}{R^{lab}} ,
\end{equation}
where $R^{lab}(T_{9})$ is obtained from $\sigma^{lab}$ measured/evaluated
in the laboratory system is the same as $R^{g.s.}$ because nuclei are in their ground states at normal conditions. The stellar enhancement factors are theoretical since no experimental data are available.  
The Eq.~\ref{myeq.SEF} shows ENDF reaction rates could be easily converted for stellar calculations with SEF.

\subsubsection{{\it kT}=8 keV}
Astrophysical reaction rates for fission and capture are shown in Tables~\ref{RR18-8},~\ref{RR102-8}. 
\subsubsection{{\it kT}=25 keV}
Astrophysical reaction rates for fission and capture are shown in Tables~\ref{RR18-25},~\ref{RR102-25}. 
\subsubsection{{\it kT}=30 keV}
Astrophysical reaction rates for fission and capture are shown in Tables~\ref{RR18-30},~\ref{RR102-30}. 
\subsubsection{{\it kT}=90 keV}
Astrophysical reaction rates for fission and capture are shown in Tables~\ref{RR18-90},~\ref{RR102-90}. 
\subsubsection{{\it kT}=1420 keV}
Astrophysical reaction rates for fission and capture are shown in Tables~\ref{RR18-1420},~\ref{RR102-1420}. 

\subsection{$r$-Process Abundances}
\begin{figure}
\begin{center}
 \includegraphics[width=.65\linewidth]{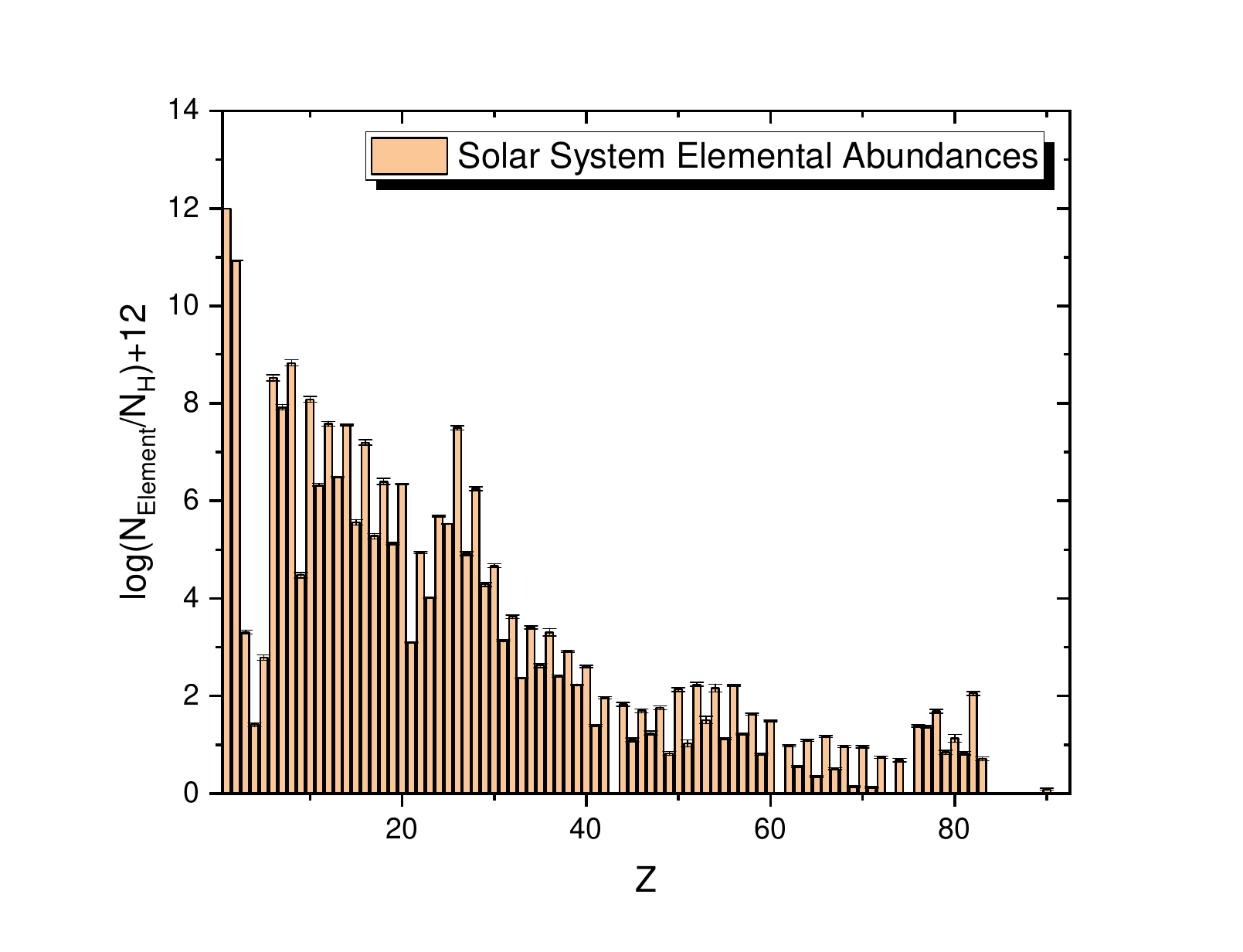}
\end{center}
\caption{ Chemical composition (Elemental abundances) of the solar system~\cite{98Gre}. 
} \label{fig:solar}
\end{figure}

The origin of chemical elements and isotope abundances is a fascinating problem.  
The observed solar system abundances are shown in Fig.~\ref{fig:solar}.  These chemical compositions were explained by explosive burning and stellar nucleosynthesis processes.  The first process produces the elements up to iron. As the coulomb barrier hinders further reactions in explosive burning, neutron capture is the only process to produce isotopes beyond $^{56}$Fe. The stellar capture processes were introduced by Burbidge, Burbidge, Fowler, Hoyle (BBFH), and Cameron~\cite{57Bur,57Cam} to explain the observed abundances of heavier than iron nuclei. 


 In this work, the classical model of stellar nucleosynthesis \cite{74Cla,82Kap} will be used to quantify the $s$-process (slow neutron capture) abundance contributions.  The classical model is based on a phenomenological and site-independent approach, and it 
assumes that the seeds for neutron captures are made entirely of $^{56}$Fe.  
The $s$-process abundance of an isotope $N_{(A)}$ depends on its precursor $N_{(A-1)}$ quantity as in
\begin{equation}\label{eq:class1}
\frac{dN_{(A)}}{dt} = \lambda_{n (A-1)} N_{(A-1)} - \big[ \lambda_{n (A)} + \lambda_{\beta (A)} \big] N_{(A)},
\end{equation}
where $\lambda_{n}$ is the neutron capture rate, and $\lambda_{\beta} = \frac{ln 2}{T_{1/2}}$ is the $\beta$-decay rate for radioactive nuclei. 
Theorizing that the temperature and neutron density are constant,  and ignoring  $s$-process branchings, the previous formula simplifies to 
\begin{equation}\label{eq:class2}
\frac{dN_{(A)}}{dt} = \sigma_{(A-1)} N_{(A-1)} + \sigma_{(A)} N_{(A)}.
\end{equation}

Equation  \ref{eq:class2} was solved analytically  for an exponential average flow of neutron exposure 
expecting that temperature remains constant over the whole timescale of the $s$-process \citep{74Cla,82Kap}. The product of MACS and isotopic abundance ($\sigma_{(A)} N_{(A)}$) was written as
\begin{equation}\label{eq:class3}
\sigma_{(A)} N_{(A)} = \frac{f N_{56}}{\tau_{0}} \prod_{i=56}^{A} \big[ 1+ \frac{1}{\sigma(i) \tau_{0}} \big]^{-1},
\end{equation}
where  $f$ and $\tau_{0}$ are the neutron fluence distribution parameters, and $N_{56}$ is the initial abundance of $^{56}$Fe seed.

The neutron capture cross section and half-life values control {\it s-}process nucleosynthesis. The neutron cross sections are small for closed neutron shell nuclei ($N\sim$ 50, 82, 126), and the process is in equilibrium in between neutron magic numbers. 
Here, the product of neutron-capture cross section (at $kT$ in mb) times solar system abundances (relative to Si = 10$^6$) as a function of atomic 
mass should be constant for equilibrium nuclei ($\sigma N$ product systematic values)~\cite{88Rol}
\begin{equation} 
\label{myeq.eq} 
\sigma_{A}N_{A}= \sigma_{A-1}N_{A-1} = constant.
\end{equation}

The Eq.~\ref{myeq.eq}  can be naively interpreted from the neutron transmission theory, where the probability of a neutron disappearance from a beam ($P_{abs}$) after traversing a thin slice of material of thickness $x$ is expressed as
\begin{eqnarray} 
\label{myeq.eq2} 
P_{abs} = exp(-\mu x) &\approx& (1 - exp(-\mu x)) \sim  \mu x \\ \nonumber
              &\approx&  (1 - exp(-\frac{\rho N_{Av}}{A} \sigma x)) \sim \frac{\rho N_{Av}}{A} \sigma x, 
\end{eqnarray}
where $\mu$, $\rho$, and $N_{Av}$ are absorption coefficient,  density, and Avogadro's number, respectively. The equation~\ref{myeq.eq2} implies that in stellar equilibrium equal numbers of neutrons are absorbed in consecutive slices of $s$-process materials.

Analysis of solar system samples shows that $s$-process abundances originate from a superposition of the
two major exponential distributions of time-integrated neutron exposure: the weak component (responsible for the production of 70 $\leq$ A $\leq$ 90 nuclei), 
and the main component (for 90 $\leq$ A $\leq$ 204 nuclei).   Earlier,  $s$-process experimental cross sections have been analyzed and fitted from $^{56}$Fe to $^{210}$Po  
as a sum of the two components individually described by Eq. \ref{eq:class3} of Ref. \citep{82Kap}. In the earlier work cross sections and solar system abundances were taken from the presently-outdated compilations \citep{82Kap,81Cam} and optimized for $s$-process only target nuclei.    Unfortunately, the two-component fitting using the present-day cross-sections and abundances was not successful~\cite{21Pri}. This outcome is consistent with the latest findings of F. K{\"a}ppeler et al. \cite{11Kap},  who indicate that the classical analysis does not firmly describe the weak component because of a limited number of $s$-process only medium nuclei and lack of equilibrium conditions. While, the empirical $\sigma N$ values for heavy nuclei not affected by branchings, are reproduced with a mean square deviation of only 3$\%$. 

Therefore, the main component only is calculated in the present work for $kT$=30 keV using evaluated cross sections, and solar system abundances of Lodders$\&$Palme~\citep{09Lod}.   Neutron fluence parameters for $s$-process  only isotopes were derived using  Eq. \ref{eq:class3}, and the derived parameters were optimized using least squares procedures. The resulting neutron fluence parameters  $f$ and $\tau_{0}$  are shown in Table \ref{tab:sfits}.
\begin{table}[h!]em abundances 
\begin{center}
 \caption{Main $s$-process neutron fluence parameters for ENDF/B-VIII.1~\cite{25Nob}, JEFF-3.3~\cite{20Plo}  and JENDL-5.0~\cite{23Iwa} libraries.  \label{tab:sfits} 
 }  

\begin{tabular}{l|ccc}
     \hline\hline
Parameters	&	ENDF/B-VIII.1  & JEFF-3.3  & JENDL-5.0 \\  \hline
f		&		0.000472$^{+0.000082}_{-0.000034}$	&  0.000422$^{+0.000073}_{-0.000020}$  &	0.000429$^{+0.000074}_{-0.000021}$  \\
$\tau_{0}$	&		0.308$^{+0.007}_{-0.020}$	& 0.345$^{+0.008}_{-0.022}$  &   0.313$^{+0.007}_{-0.019}$   \\    \hline\hline
  \end{tabular}
\end{center}
\end{table}
 
\begin{figure}
\begin{center}
\includegraphics[width=.65\linewidth]{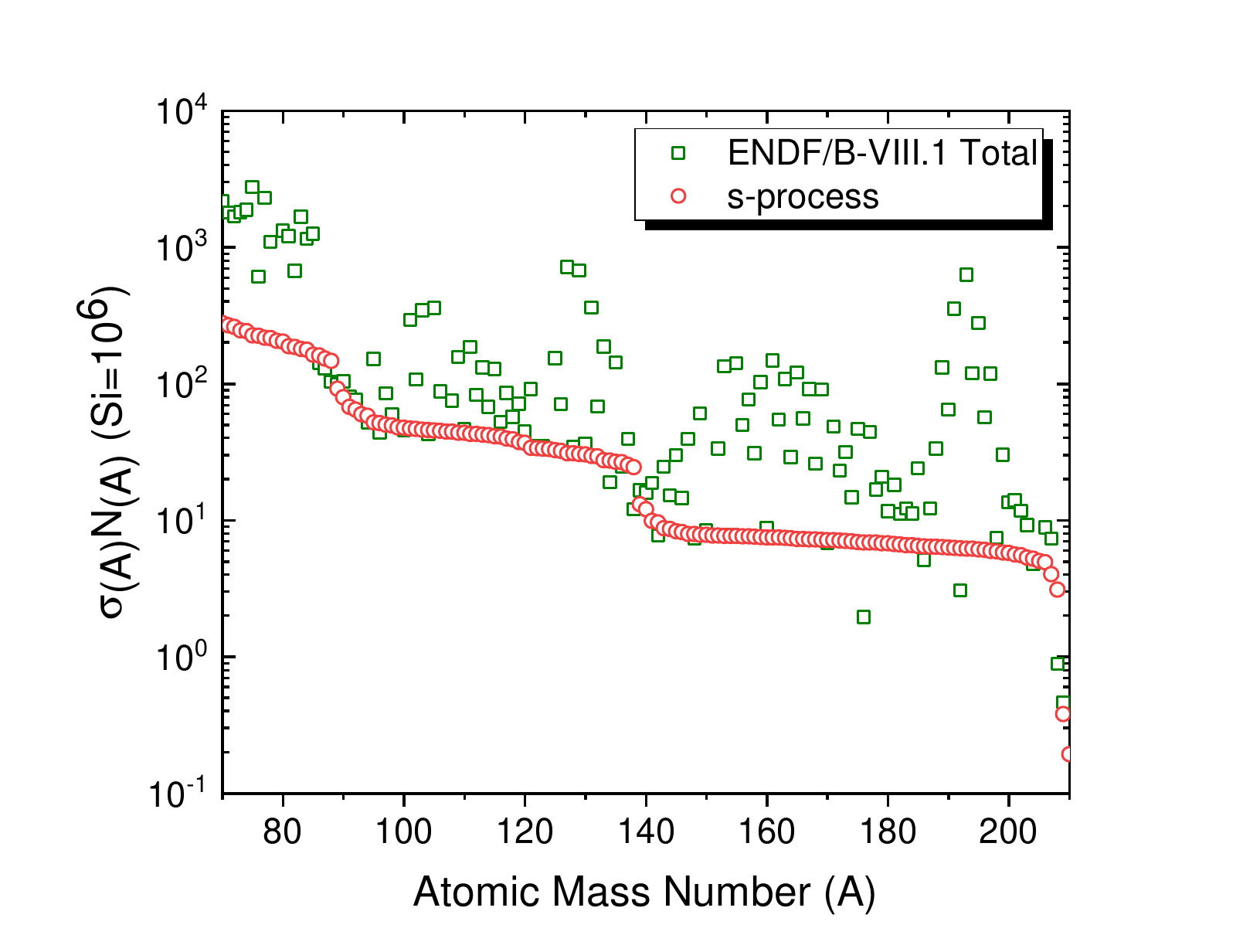}
\end{center}
\caption{ 
ENDF/B-VIII.1 (n,$\gamma$) MACS at 30 keV times solar system abundances (circles) as a function of the atomic mass number for $s$- and $r$-process nuclides and $\sigma$N$_{(A)}$ values from a classical $s$-process calculation.  
\label{fig:snuclei}}
\end{figure}

Accordingly, the $s$-process contribution to solar system abundances can be estimated using  neutron fluence parameters and compared with 
 observed values. The ENDF/B-VIII.1 MACS at $kT$=30 keV times solar abundance and expected classical $s$-process model product values are shown in Fig. \ref{fig:snuclei}.  In the Figure, we depict MACS calculated from ENDF/B-VIII.1 library times solar abundances for all stable isotopes as the ENDF/B-VIII.1 Total and compare them with the $s$-process expected values~\cite{21Pri}. 

\begin{figure}
\begin{center}
 \includegraphics[width=.65\linewidth]{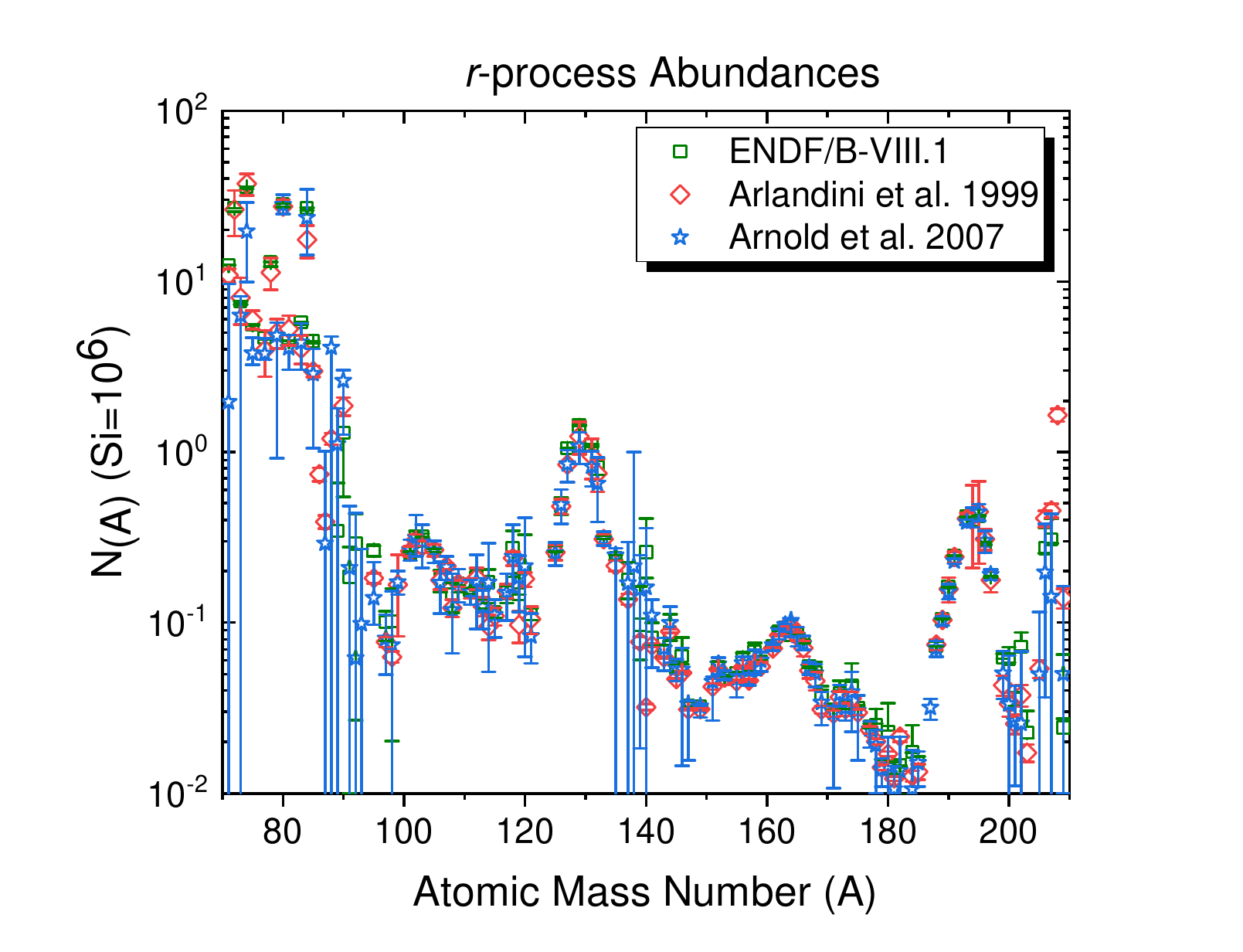}
\end{center}
\caption{ 
 Solar \mbox{$r$-process} abundances for nuclides that are produced by both the $s$- and \mbox{$r$-processes} derived from ENDF/B-VIII.1 (squares) compared with those obtained by Arlandini {\it et al.}~\cite{99Arl} (diamonds) and Arnould {\it et al.}~\cite{99Gor,07Arn} (stars). 
} \label{fig:rnuclei2}
\end{figure}

\begin{figure}
\begin{center}
 \includegraphics[width=.65\linewidth]{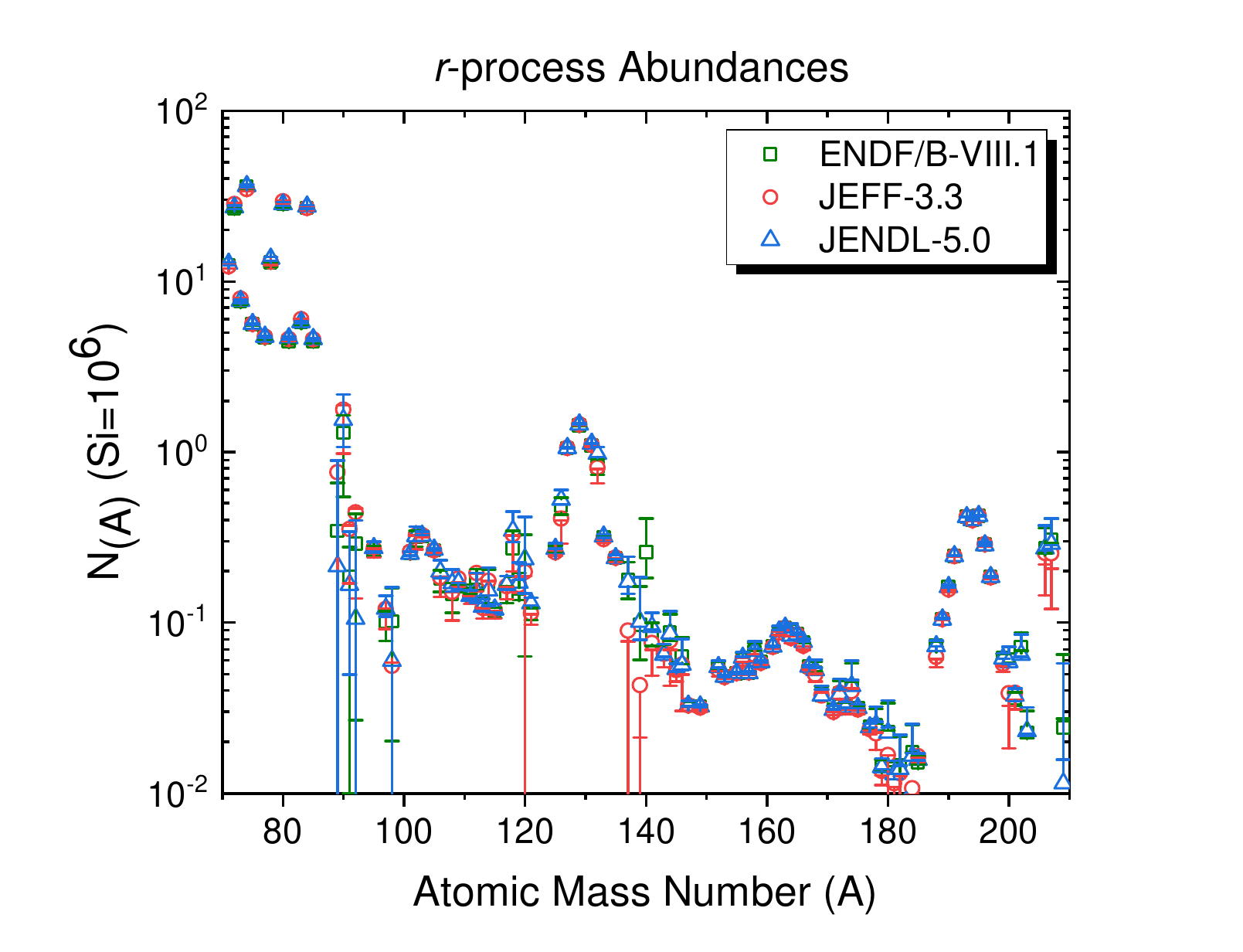}
\end{center}
\caption{ 
 Solar \mbox{$r$-process} abundances for nuclides that are produced by both the $s$- and \mbox{$r$-processes} derived from ENDF/B-VIII.1 (squares),  JEFF-3.3 (circles) and JENDL-5.0  (triangles) libraries. 
} \label{fig:rnuclei3}
\end{figure}

The data in the figure indicate a surplus production for many nuclei compared with the $s$-process systematic values. 
The Fig. \ref{fig:snuclei} surplus is commonly attributed to the $r$-process (rapid neutron capture) contribution, and it can be derived by subtracting the 
expected classical model $s$-process production from the total ENDF/B values and divided by cross sections. The shown in Fig.~\ref{fig:rnuclei2} ENDF/B-VIII.1 $r$-process abundances are in agreement with Arlandini  {\it  et al.}~\cite{99Arl}  and Arnould  {\it  et al.}~\cite{99Gor,07Arn} values. Furthermore, Fig.~\ref{fig:rnuclei3}  shows that JEFF-3.3 and JENDL-5.0 produce good results. In our previous work, we observed issues with $^{138}$Ba and $^{140}$Ce cross sections in ENDF/B-VIII.0 library~\cite{21Pri}. Complementary analysis of the current work $r$-process abundances shows that the $^{140}$Ce cross section issue is resolved in the ENDF/B-VIIII.1 library, yet $^{138}$Ba still needs re-evaluation and JEFF-3.3 and JENDL-5.0 underestimate resonance region cross sections for both nuclei.  Numerical values for solar system $r$-process abundances calculated with the ENDF/B-VIII.1, JEFF-3.3, and JENDL-5.0 libraries are given in Table~\ref{RP-30}.

 \subsubsection{Nuclear Structure and Stellar Abundances}
 The observation of the second $N$=82 and the third $N$=126 $r$-process peaks in Figs.~\ref{fig:rnuclei2}-\ref{fig:rnuclei3} provides strong evidence of the existence of neutron magic numbers beyond the valley of stability. At the same time, the lack of the first $r$-process peak may indicate that $N$=50 is not a good quantum (magic) number for neutron-rich nuclei or neutron reaction flow changes for medium nuclei. The observed synergy between nuclear astrophysics and nuclear structure physics requires additional research and close cooperation between the two communities~\cite{21Pri}.

\subsection{Summary of the ENDF Integral Values}
The current work is the complete calculation of ENDF integral values using the latest libraries, and it supersedes the previous works~\cite{05Nak,10Pri,12Pri,21Pri}. The list of major features includes:
\begin{itemize}
\item An extended list of integral values: neutron thermal cross sections, Westcott factors, resonance integrals, Maxwelian-averaged cross sections for $kT$=8, 25, 30, 90, and 1420 keV, and astrophysical reaction rates
\item Improved numerical integration and T=0, 293.6$^{\circ}$ K with 0.1$\%$ Doppler broadening
\item Calculation of uncertainties using ENDF libraries covariances
\item More than 99$\%$ coverage of the $s$-process path (the $s$-process path depends on stellar conditions and may slightly deviate from the standard scenario)
\item Unified neutron material grid for evaluated neutron libraries
\item Coverage of the five major libraries
\end{itemize}

\subsubsection{False Positive Comparisons}

It is a frequent practice in ENDF evaluations to adopt (borrow) or slightly correct high-quality evaluations from other libraries. Some libraries follow this practice more often than others.  As a result, some ENDF integral values look identical as in $^{240-250}$Cm and create an illusion of a perfect agreement between different libraries. ENDF curium data evaluations originate from distinct versions of the JENDL libraries, and numerical coincidences do not provide validation. Therefore,  the observed agreements require additional research and consultation with other ENDF and EXFOR libraries.

\section{Conclusion and Outlook}
\label{sec:conclusion}
Worldwide demands for the development of new nuclear energy and astrophysics applications provided a strong motivation for this work. 
A complete calculation of  Westcott factors, resonance integrals,  Maxwellian-averaged cross sections, astrophysical reaction rates, and their uncertainties has been performed. Neutron thermal cross section data were extracted from the Doppler broaden ENDF libraries. 
Present data were analyzed using benchmarks, where available. Data analysis indicates substantial progress in 
nuclear data libraries' quality and their importance for multiple applications. 
\section*{Acknowledgments}

The author is indebted to Dr. Alberto Mengoni (ENEA, Bologna $\&$ INFN, Sezione di Bologna) and Dr. Pavel Oblo{\v z}insk{\' y} (BNL) for suggesting nuclear astrophysics calculations with ENDF libraries, and Drs. Viktor Zerkin (IAEA), Marco Pigni (ORNL), Alejandro Sonzogni, and the late  Said F. Mughaghab (BNL) for help with ENDF file Doppler broadening, covariances, and helpful discussions, respectively.  Productive interactions with the IAEA Technical Meeting (TM) members on Thermal Capture and Prompt Capture Gamma Databases participants are also acknowledged.  Work at Brookhaven was funded by the Office of Nuclear Physics, Office of Science of the U.S. Department
of Energy, under Contract No. DE-SC0012704 with Brookhaven Science Associates, LLC.  \\ \\

\appendix
\section{Appendix}
\datatables 


\end{center}
\normalsize

\clearpage


\begin{thebibliography}{999}

\bibitem{11Chad} M.B.~Chadwick, M.W.~Herman, P.~Oblo{\v z}insk{\' y}, M.E.~Dunn, Y.~Danon,  A.C.~Kahler, D.L.~Smith,
    B.~Pritychenko, G.~Arbanas, R.~Arcilla, R.~Brewer, D.A.~Brown, R.~Capote,  A.D.~Carlson, Y.S.~Cho, H.~Derrien,  K.~Guber,
    G.M.~Hale, S.~Hoblit,  S.~Holloway, T.D.~Johnson, T.~Kawano,  B.C.~Kiedrowski,  H.~Kim, S.~Kunieda, N.M.~Larson, L.~Leal,
    J.P.~Lestone,  R.C.~Little, E.A.~McCutchan, R.E.~MacFarlane, M.~MacInnes, C.M.~Mattoon, R.D.~McKnight, S.F.~Mughabghab,
    G.P.A.~Nobre, G.~Palmiotti, A.~Palumbo, M.T.~Pigni, V.G.~Pronyaev,  R.O.~Sayer, A.A.~Sonzogni, N.C.~Summers, P.~Talou,
    I.J.~Thompson, A.~Trkov, R.L.~Vogt, S.C.~van der Marck,  A.~Wallner, M.C.~White, D.~Wiarda, P.G.~Young, ``ENDF/B-VII.1 Nuclear Data for Science and Technology: Cross Sections, Covariances, Fission Product Yields and Decay Data," {\sc  Nucl. Data Sheets} {\bf 112}, 2887 (2011).

\bibitem{06Mugh} S.F.~Mughabghab, {\sc Atlas of Neutron Resonances, Resonance Parameters and Neutron Cross Sections Z = 1-100}, Elsevier (2006).
\bibitem{18Mugh}  S.F.~Mughabghab, {\sc Atlas of Neutron Resonances, Resonance Parameters and Thermal Cross Sections Z = 1-60, 61-102,} {\bf 1,2}, Elsevier (2018). 

\bibitem{05Nak} T.~Nakagawa, S.~Chiba, T.~Hayakawa, T.~Kajino, ``Maxwellian-averaged Neutron-induced Reaction Cross Sections and Astrophysical Reaction Rates for kT = 1 keV to 1 MeV Calculated from Microscopic Neutron Cross Section Library JENDL-3.3,"  {\sc At. Data and Nucl. Data Tables} {\bf 91}, 77 (2005).
\bibitem{10Pri} B.~Pritychenko, S.F.~Mughabghab, A.A.~Sonzogni, ``Calculations of Maxwellian-averaged Cross Sections and Astrophysical Reaction Rates Using the ENDF/B-VII.0, JEFF-3.1, JENDL-3.3 and ENDF/B-VI.8 Evaluated Nuclear Reaction Data Libraries," {\sc At. Data and Nucl. Data Tables} {\bf 96}, 645 (2010).
\bibitem{12Pri} B.~Pritychenko, S.F.~Mughabghab, ``Neutron Thermal Cross Sections, Westcott Factors, Resonance Integrals, Maxwellian Averaged Cross Sections and Astrophysical Reaction Rates Calculated from the ENDF/B-VII.1, JEFF-3.1.2, JENDL-4.0, ROSFOND-2010, CENDL-3.1 and EAF-2010 Evaluated Data Libraries," {\sc Nucl. Data Sheets} {\bf 113}, 3120 (2012).
\bibitem{21Pri} B.~Pritychenko, ``Capitalizing on nuclear data libraries' comprehensiveness to obtain solar system $r$-process abundances," {\sc J. Phys. G} {\bf 48}, 08LT01 (2021).

\bibitem{25Nob} G.P.A.~Nobre,  R.~Capote, M.T.~Pigni,  A.~Trkov, C.M.~Mattoon,  D.~Neudecker, D.A.~Brown,1 M.B.~Chadwick, A.C.~Kahler, N.A.~Kleedtke, M.~Zerkle, A.I.~Hawari, C.W.~Chapman, N.C.~Fleming, J.L.~Wormald, K.~Rami{\' c}, Y.~Danon, N.A.~Gibson, P.~Brain, M.W.~Paris, G.M.~Hale, I.J.~Thompson, D.P.~Barry, I.~Stetcu, W.~Haeck, A.E.~Lovell, M.R.~Mumpower, G.~Potel, K.~Kravvaris, G.~Noguere, J.D.~McDonnell, A.D.~Carlson, M.~Dunn, T.~Kawano, D.~Wiarda, I.~Al-Qasir, G.~Arbanas, R.~Arcilla, B.~Beck, D.~Bernard, R.~Beyer, J.M.~Brown, O.~Cabellos, R.J.~Casperson, Y.~Cheng, E.V.~Chimanski, R.~Coles, M.~Cornock, J.~Cotchen, J.P.W.~Crozier, D.E.~Cullen, A.~Daskalakis, M.-A.~Descalle, D.D.~DiJulio, P.~Dimitriou, A.C.~Dreyfuss, I.~Duran, R.~Ferrer, T.~Gaines, V.~Gillette,
G.~Gert, K.H.~Guber, J.D.~Haverkamp, M.W.~Herman, J.~Holmes, M.~Hursin, N.~Jisrawi, A.R.~Junghans, K.J.~Kelly, H.I.~Kim, K.S.~Kim, A.J.~Koning, M.~Ko{\" s}t{\' a}l, B.K.~Laramee,A.~Lauer-Coles, L.~Leal, H.Y.~Lee, A.M.~Lewis, J.~Malec, J.I.~M{\' a}rquez Dami{\' a}n, W.J.~Marshall, A.~Mattera, G.~Muhrer, A.~Ney, W.E.~Ormand, D.K.~Parsons, C.M.~Percher, B.~Pritychenko, V.G.~Pronyaev, A.~Qteish, S.~Quaglioni, M.~Rapp, J.J.~Ressler, M.~Rising, D.~Rochman, P.K.~Romano, D.~Roubtsov, G.~Schnabel, M.~Schulc, G.J.~Siemers, A.A.~Sonzogni, P.~Talou, J.~Thompson, T.H.~Trumbull, S.C.~van der Marck, M.~Vorabbi, C.~Wemple, K.A.~Wendt, M.~White, and R.Q.~Wright,  ``ENDF/B-VIII.1: Updated Nuclear Reaction Data Library for Science and Applications," {\sc  Nucl. Data Sheets} {\bf X}, Y (2025). 

\bibitem{20Plo} A.J.M.~Plompen, O.~Cabellos, C.~De Saint Jean, M.~Fleming, A.~Algora, M.~Angelone, P.~Archier, E.~Bauge, O.~Bersillon, A.~Blokhin, F.~Cantargi, A.~Chebboubi, C.~Diez, H.~Duarte, E.~Dupont, J.~Dyrda, B.~Erasmus, L.~Fiorito, U.~Fischer, D.~Flammini, D.~Foligno, M.R.~Gilbert, J.R.~Granada, W.~Haeck, F.-J.~Hambsch, P.~Helgesson, S.~Hilaire, I.~Hill, M.~Hursin, R.~Ichou, R.~Jacqmin, B.~Jansky, C.~Jouanne, M.A.~Kellett, D.H.~Kim, H.I.~Kim, I.~Kodeli, A.J.~Koning, A.Yu.~Konobeyev, S.~Kopecky, B.~Kos, A.~Kr{\' a}sa, L.C.~Leal, N.~Leclaire, P.~Leconte, Y.O.~Lee, H.~Leeb, O.~Litaize, M.~Majerle, J.I.~M{\' a}rquez Dami{ a}n, F.~Michel-Sendis, R.W.~Mills, B.~Morillon, G.~Nogu{\'  e}re, M.~Pecchia, S.~Pelloni, P.~Pereslavtsev, R.J.~Perry, D.~Rochman, A.~R{\" o}hrmoser, P.~Romain, P.~Romojaro, D.~Roubtsov, P.~Sauvan, P.~Schillebeeckx, K.H.~Schmidt, O.~Serot, S.~Simakov, I.~Sirakov, H.~Sj{\" o}strand, A.~Stankovskiy, J.C.~Sublet, P.~Tamagno, A.~Trkov, S.~van der Marck, F.~{\' A}lvarez-Velarde, R.~Villari, T.C.~Ware, K.~Yokoyama, G.~{\' Z}erovnik  ``The joint evaluated fission and fusion nuclear data library, JEFF-3.3," {\sc Eur. Phys. J.} {\bf A 56}, 181 (2020).
\bibitem{23Iwa} O.~Iwamoto, N.~Iwamoto, S.~Kunieda, F.~Minato, Sh.~Nakayama,
Yu.~Abe, K.~Tsubakihara, Sh.~Okumura, Ch.~Ishizuka, T.~Yoshida,
S.~Chiba, N.~Otuka, J.-Ch.~Sublet, H.~Iwamoto, K.~Yamamoto,
Y.~Nagaya, K.~Tada, Ch.~Konno, N.~Matsuda, K.~Yokoyama, H.~Taninaka,
A.~Oizumi, M.~Fukushima, Sh.~Okita, G.~Chiba, S.~Sato, M.~Ohta, S.~Kwon ``Japanese evaluated nuclear data library version 5: JENDL-5," {\sc J. Nucl. Sci. Tech.} {\bf 60}, 1 (2023).
\bibitem{20Ge} Zh.~Ge, R.Xu, H.~Wu, Yu.~Zhang, G.~Chen, Y.~Jin, N.~Shu, Y.~Chen, X.~Tao, Yu.~Tian, P.~Liu, J.~Qian, J.~Wang, H.~Zhang, L.~Liu, X.Huang, ``CENDL-3.2: The new version of Chinese general purpose evaluated nuclear data library," {\sc EPJ Web of Conferences} {\bf 239}, 09001 (2020).
\bibitem{16Blo} A.I.~Blokhin, E.V.~Gai, A.V.~Ignatyuk, I.I.~Koba, V.N.~Manokhin, V.G.~Pronyaev, ``New Version of Neutron Evaluated Data Library BROND-3.1," {\sc Neutr. Const. Param.} {\bf 2},  2:5 (2016).

\bibitem{12Kon} A.J.~Koning, D.~Rochman, ``Modern Nuclear Data Evaluation with the TALYS Code System," {\sc Nucl. Data Sheets} {\bf 113}, 2841 (2012).
\bibitem{20Pri} B.~Pritychenko, ``Determination of Solar System R-Process Abundances using ENDF/B-VIII.0 and TENDL-2015 libraries," Brookhaven National Laboratory Report BNL-220698-2020-INRE (2020);   arXiv:2012.06728v2 [astro-ph.SR]  (2021).
\bibitem{25Roc} D.~Rochman, A.~Koning, S.~Goriely, S.~Hilaire, ``TENDL-astro: A new nuclear data set for astrophysics interest," {\sc Nucl. Phys.} {\bf A1053}, 122951 (2025).

\bibitem{X4} N.~Otuka, E.~Dupont, V.~Semkova, B.~Pritychenko, A.I.~Blokhin, M.~Aikawa, S.~Babykina,
M.~Bossant, G.~Chen, S.~Dunaeva, R.A.~Forrest, T.~Fukahori, N.~Furutachi, S.~Ganesan, Z.~Ge, 
O.~Gritzay, M.~Herman, S.~Hlavac, K.~Kato, B.~Lalremruata, Y.O.~Lee, A.~Makinaga,
K.~Matsumoto, M.~Mikhaylyukova, G.~Pikulina, V.G.~Pronyaev, A.~Saxena, O.~Schwerer,
S.P.~Simakov, N.~Soppera, R.~Suzuki, S.~Takacs, X.~Tao, S.~Taova, F.~Tarkanyi, V.V.~Varlamov,
J.~Wang, S.C.~Yang, V.~Zerkin, Y.~Zhuang, ``Towards a More Complete and Accurate Experimental
Nuclear Reaction Data Library (EXFOR): International Collaboration between Nuclear Reaction Data
Centres (NRDC)," {\sc Nucl. Data Sheets} {\bf 120} 272, (2014).
\bibitem{EXFOR}  V.V.~Zerkin, B.~Pritychenko, ``The experimental nuclear reaction data (EXFOR): Extended computer database and Web retrieval system," {\sc Nucl. Instr. Meth. Phys. Res.} {\bf A 888}, 31 (2018). 

\bibitem{24Neu} D.~Neudecker, K.J.~Kelly, A.D.~Carlson, B.~Pritychenko, M.J.~Grosskopf, S.A.~Vander Wiel, R.C.~Haight, D.A.~Brown, ``Re-evaluating the Prompt Fission Neutron Spectrum of Spontaneously Fissioning $^{252}$Cf," Los Alamos National Laboratory Report LA-UR-24-24178 (2024).
\bibitem{23Ves} D.~Vescovi, R.~Reifarth, E.~Lorenz, A.~Elbe, ``The ASTRAL database for neutron-capture nucleosynthesis studies," {\sc EPJ Web of Conf.} {\bf 279}, 11011 (2023).
\bibitem{09Car} A.D.~ Carlson, V.G.~ Pronyaev, D.L.~ Smith, N.M.~ Larson, Zh.~ Chen, G.M.~ Hale, F.-J.~ Hambsch, E.V.~ Gai, S.-Y.~ Oh, S.A.~ Badikov, T.~ Kawano, H.M.~ Hofmann, H.~ Vonach, S.~ Tagesen,  ``International Evaluation of Neutron Cross Section Standards," {\sc Nucl. Data Sheets} {\bf 110} 3215 (2009). 
\bibitem{15Car} A.D.~ Carlson, V.G.~ Pronyaev, R.~ Capote, G.M.~ Hale, F.-J.~ Hambsch, T.~Kawano, S.~ Kuneida, W.~ Mannhart,  R.O.~ Nelson, D.~ Neudecker, P.~ Schillebeeckx, S.~Simakov,  D.L.~ Smith, P.~Talou, X.~Tao,  A.~ Wallner, W.~ Wang,   ``Recent Work Leading Towards a New Evaluation of the Neutron Standards," {\sc Nucl. Data Sheets} {\bf 123}, 27 (2015).
\bibitem{18Car} A.D.~ Carlson, V.G.~ Pronyaev, R.~ Capote, G.M.~ Hale, Z.-P.~ Chen,  I.~ Duran, F.-J.~ Hambsch, S.~ Kuneida, W.~ Mannhart, B.~ Marcinkevicius, R.O.~ Nelson, D.~ Neudecker, G.~ Noguere, M.~ Paris, S.P.~ Simakov, P.~ Schillebeeckx, D.L.~ Smith, X.~ Tao, A.~ Trkov, A.~ Wallner, W.~ Wang,``Evaluation of the Neutron Data Standards,"  {\sc Nucl. Data Sheets} {\bf 148}, 143 (2018).
\bibitem{20Tr}  A.~Trkov, P.J.~Griffin, S.P.~Simakov, L.R.~Greenwood, K.I.~Zolotarev, R.~Capote, D.L.~Aldama, V.~Chechev, C.~Destouches, A.C.~Kahler, C.~Konno, M.~Kostal, M.~Majerle, E.~Malambu, M.~Ohta, V.G.~Pronyaev, V.~Radulovic, S.~Sato, M.~Schulc, E.~Simeckova, I.~Vavtar, J.~Wagemans, M.~White, H.~Yashima, ``IRDFF-II: A New Neutron Metrology Library," {\sc Nucl.Data Sheets} {\bf 163}, 1 (2020). 

\bibitem{10Phy} PHYSOR 2010, ``Advances in Reactor Physics to Power Nuclear Renaissance", Pittsburgh, PA, May 9-14 (2010). Available from $\langle$http://physor2010.org$\rangle$.
\bibitem{11He} A.~Trkov, M.~Herman, D. A.~Brown,  ``ENDF-6 Formats Manual: Data Formats and Procedures for the Evaluated Nuclear Data File ENDF/B-VI and ENDF/B-VII",  Brookhaven National Laboratory Report BNL-90365-2009 Rev.2, CSEWG Document ENDF-102, December 2011.
\bibitem{CSEWG} {\sc Cross Section Evaluation Working Group (CSEWG)}, Available from $\langle$http://www.nndc.bnl.gov/csewg$\rangle$.
\bibitem{USNDP} {\sc United States Nuclear Data Program (USNDP)}, Available from $\langle$http://www.nndc.bnl.gov/usndp$\rangle$.
\bibitem{06Pri} B.~Pritychenko, A.A.~Sonzogni, D.F.~Winchell, V.V.~Zerkin,
R.~Arcilla, T.W.~Burrows, C.L.~Dunford, M.W.~Herman, V.~McLane, P.~Oblo{\v z}insk{\' y},  Y.~Sanborn, J.K.~Tuli, ``Nuclear Reaction and Structure Data Services of the National Nuclear Data Center,"  {\sc Ann. Nucl. Energy} {\bf 33}, 390 (2006).
\bibitem{WPEC} {\sc Working Party on International Nuclear Data Evaluation Co-operation (WPEC)}, Available from $\langle$http://www.nea.fr/html/science/wpec/$\rangle$.
\bibitem{04Hum} D.P.~Humbert, A.L.~Nichols, O.~Schwerer, ``IAEA Nuclear Data Section: provision of atomic and nuclear databases for user applications," {\sc Appl. Rad. Isot.} {\bf 60}, 311 (2004).
\bibitem{06Chad} M.B. Chadwick, P. Oblo{\v z}insk{\' y}, M.W. Herman, N.M.~Greene, R.D.~McKnight, D.L.~Smith, P.G.~Young, R.E.~MacFarlane, G.M.~Hale, S.C.~Frankle, A.C.~Kahler, T.~Kawano, R.C.~Little, D.G.~Madland, P.~Moller, R.D.~Mosteller, P.R.~Page, P.~Talou, H.~Trellue, M.C.~White, W.B.~Wilson, R.~Arcilla, C.L.~Dunford, S.F.~Mughabghab, B.~Pritychenko, D.~Rochman, A.A.~Sonzogni, C.R.~Lubitz, T.H.~Trumbull, J.P.~Weinman, D.A.~Brown, D.E.~Cullen, D.P.~Heinrichs, D.P.~McNabb, H.~Derrien, M.E.~Dunn, N.M.~Larson, L.C.~Leal, A.D.~Carlson, R.C.~Block, J.B.~Briggs, E.T.~Cheng, H.C.~Huria, M.L.~Zerkle, K.S.~Kozier, A.~Courcelle, V.~Pronyaev, S.C.~van der Marck,  ``ENDF/B-VII.0: Next Generation Evaluated Nuclear Data Library for Nuclear Science and Technology," {\sc  Nucl. Data Sheets} {\bf 107}, 2931 (2006).
\bibitem{01Cse} CSEWG-Collaboration, ``Evaluated Nuclear Data File ENDF/B-VI.8," Available from $\langle$http://www.nndc.bnl.gov/endf/$\rangle$.
\bibitem{11Kon} A.J.~Koning, E.~Bauge, C.J.~Dean, E.~Dupont, U.~Fischer, R.A.~Forrest, R.~Jacqmin, H.~Leeb, M.A.~Kellett, R.W.~Mills, C.~Nordborg, M.~Pescarini, Y.~Rugama, P.~Rullhusen, ``Status of the JEFF Nuclear Data Library," {\sc J. of the Korean Phys. Soc.}  {\bf 59}, No. 2, 1057 (2011).
\bibitem{02Shi} K.~Shibata, T.~Kawano, T.~Nakagawa, O.~Iwamoto, J.~Katakura, T.~Fukahori, S.~Chiba, A.~Hasegawa, T.~Murata, H.~Matsunobu, T.~Ohsawa, Y.~Nakajima, T.~Yoshida, A.~Zukeran, M.~Kawai, M.~Baba, M.~Ishikawa, T.~Asami, T.~Watanabe, Y.~Watanabe, M.~Igashira, N.~Yamamuro, H.~Kitazawa, N.~Yamano, H.~Takano, ``Japanese Evaluated Nuclear Data Library Version 3 Revision-3: JENDL-3.3," {\sc Nucl. Sci. Tech.} {\bf 39}, 1125 (2002). 
\bibitem{11Shi} K.~Shibata, T.~Kawano, T.~Nakagawa, N.~Iwamoto, A.~Ichihara, S.~Kunieda, S.~Chiba, K.~Furutaka, N.~Otuka, T.~Ohsawa, T.~Murata, H.~Matsunobu, A.~Zukeran, S.~Kamada, J.~Katakura,  ``JENDL-4.0: A New Library for Nuclear Science and Engineering," {\sc Nucl. Sci. Tech.} {\bf 48}, 1 (2011). 
\bibitem{11Zhi} Z.G.~Ge, Z.X.~Zhao, H.H.~Xia,  Y.X.~Zhuang,  T.J.~Liu,  J.S.~Zhang,  H.C.~Wu,   ``The Updated Version of Chinese Evaluated Nuclear Data Library (CENDL-3.1)," {\sc J. of the Korean Phys. Soc.}, {\bf 59}, No. 2, 1052 (2011).
\bibitem{07Zab} S.V.~Zabrodskaya, A.V.~Ignatyuk , V.N.~Koscheev, V.N.~Mahokhin,  M.N.~Nikolaev,  V.G.~Pronyaev, ``ROSFOND - Rossiyskaya Natsionalnaya Biblioteka Nejtronnykh Dannykh," {\sc VANT}, Nucl. Constants {\bf 1-2}, 3 (2007).
\bibitem{BROND} A.I.~Blokhin, B.I.~Fursov, A.V.~Ignatyuk, V.N.~Koshcheev, V.~Kulikov, B.D.~Kuzminov, V.N.~Manokhin, M.N.~Nikolaev, ``Current Status of Russian Evaluated Neutron Data Libraries," {\sc Proceedings of Int. Conf. on Nuclear Data and Technology}, Gatlinburg, Tennessee, USA, May 9-13, 1994; {\bf 2}, 695 (1994).





\bibitem{10Mac} R.E.~MacFarlane, A.C.~Kahler, ``Methods for Processing ENDF/B-VII with NJOY," {\sc Nucl. Data Sheets} {\bf 111},  2739 (2010).
\bibitem{07Cul} D.E.~Cullen, ``The ENDF/B Pre-processing codes (PREPRO),"  Available from $\langle$http://www-nds.iaea.org/ ndspub/endf/prepro/$\rangle$.
\bibitem{23Pig}  M.T.~Pigni, J.~McDonnell, J.H.~Guber, ``Resonance Parameter Evaluation of n+$^{88}$Sr reactions for ENDF/B-VIII.1 Library," Oak Ridge National Laboratory Report ORNL/LTR-2023/3004 (2023).


\bibitem{55We} C.H.~Westcott, ``The Specification of Neutron Flux and Nuclear Cross-sections in Reactor Calculations," {\sc J. Nucl. Energy} {\bf 2}, 59 (1955).
\bibitem{58We} C.H.~Westcott, W.H.~Walker, T.K.~Alexander, ``Effective Cross Sections and Cadmium Ratios for the Neutron Spectra of Thermal Reactors," Proc. 2$^{nd}$ U.N. Int. Conf. on the Peaceful Usage of Atomic Energy {\bf 15}, 202 (1958).

\bibitem{inter} O.~Ozer, {\sc Computer Code INTER}; Available from $\langle$http://www.nndc.bnl.gov/nndcscr/endf/endf-util$\rangle$.
\bibitem{99Ho} N.E.~Holden, ``Temperature Dependence of the Westcott g-factor for Neutron Reactions in Activation Analysis," {\sc Pure Appl. Chem.} {\bf 71}, 2309 (1999).
\bibitem{07Ch} H.D.~Choi, A.~Trkov, ``Database of Prompt Gamma Rays from Slow Neutron Capture for Elemental Analysis," IAEA, Vienna, 5 (2007).

\bibitem{66Lam} J.R.~Lamarsh, ``Nuclear Reactor Theory."  Addison-Wesley, pp. 234-235 (1966).
\bibitem{80Bak} V.C.~Baker, J.H.~Marable, ``Resonance Integral Calculations for Isolated Rods Containing Oxides of $^{238}$U and $^{232}$Th," Oak Ridge National Laborartory Report ORNL/TM-6376, (1980).
\bibitem{68De} D.~DeSoete, R.~Gijbels, J.~Hoste, ``Neutron, Photon and Charged Particle Reactions for Activation Analysis,"   {\sc Nat. Bureau of Standards Special Pub. 312} {\bf 2}, 699 (1969); J.R. DeVoe, P.D. LaFleur (editors), ``Modern Trends in Activation Analysis," {\sc Proc. of the 1968 International Conference}, Gaithersburg, Maryland, October 7-11 (1968). Available from $\langle$http://books.google.com$\rangle$.
\bibitem{02Ba} E.M.~Baum, H.D.~Knox, T.R.~Miller, ``Nuclides and Isotopes, Chart of Nuclides," Wall Chart Information Booklet, Sixteen Edition, KAPL (2002).
\bibitem{10BPri} B.~Pritychenko, ``Complete Calculation of Evaluated Maxwellian-averaged Cross Sections and their Uncertainties for s-process Nucleosynthesis," {\sc Proceedings of 11$^{th}$ Symposium on Nuclei in the Cosmos, NIC XI}, July 19-23 (2010), Heidelberg, Germany;  Brookhaven National Laboratory Report BNL-93963-2010-CP.
\bibitem{09Math} ``Wolfram Mathematica Online Integrator," Available from  $\langle$http://integrals.wolfram.com$\rangle$. 
\bibitem{92Bev} P.R.~Bevington, D.K.~Robinson, {\it Data Reduction and Error Analysis for the Physical Sciences}, WCB/McGraw-Hill, (1992).


\bibitem{New23}  D.~Neudecker, A.~Lewis, E.~Matthews, J.~Vanhoy, R.~Haight, D.~Smith, P.~Talou, S.~Croft, A.~Carlson, B.~Pierson, A.~Wallner, A.~Al-Adili, L.~Bernstein, R.~Capote, M.J.~Devlin, M.~Drosg, D.~Duke, S.~Finch, M.~Herman, K.~Kelly, A.~Koning, A.~Lovell, P.~Marini, K.~Montoya, G.P.A.~Nobre, M.~Paris, B.~Pritychenko, H.~Sj{\" o}strand, L.~Snyder, V.~Sobes, A.~Solders, J.~Taieb, ``Templates of expected measurement uncertainties," {\sc EPJ N} {\bf 9}, 35 (2023).
\bibitem{Lew23} A.M.~Lewis, D.~Neudecker, A.D.~Carlson, D.L.~Smith, I.~Thompson, A.~Wallner, D.P.~Barry, L.A.~Bernstein, R.C.~Block, S.~Croft, Y.~Danon, M.~Drosg, R.C.~Haight, M.W.~Herman, H.Y.~Lee, N.~Otuka, H.~Sj{\" o}strand, V.~Sobes, ``Templates of expected measurement uncertainties for neutron-induced capture and charged-particle production cross section observables," {\sc EPJ N} {\bf 9},  33 (2023).
\bibitem{Van23} J.R.~Vanhoy, R.C.~Haight, S.F.~Hicks, M.~Devlin, D.~Neudecker, M.~Herman, A.~Koning, K.J.~Kelly, I.~Thompson, ``Templates of expected measurement uncertainties for (n, xn) cross sections," {\sc EPJ N} {\bf 9}, 31 (2023).
\bibitem{Le23} A.M.~Lewis, A.D.~Carlson, D.L~Smith, D.P.~Barry, R.C.~Block, S.~Croft, Y.~Danon, M.~Drosg, M.W.~Herman, D.~Neudecker, N.~Otuka, H.~Sj{\" o}strand, V.~Sobes, ``Templates of expected measurement uncertainties for total neutron cross-section observables," {\sc EPJ N} {\bf 9}, 34 (2023).
\bibitem{Cap20} R.~Capote, S.~Badikov, A.D.~Carlson, I.~Duran, F.~Gunsing, D.~Neudecker, V.G.~Pronyaev, P.~Schillebeeckx, G.~Schnabel, D.L.~Smith, A.~Wallner, ``Unrecognized Sources of Uncertainties (USU) in Experimental Nuclear Data," {\sc  Nucl. Data Sheets} {\bf 163}, 191 (2020).
\bibitem{11Kap} F.~K{\"a}ppeler, R.~Gallino, S.~Bisterzo, W.~Aoki,  ``The s process: Nuclear physics, stellar models, and observations," {\sc Rev. Mod. Phys.} {\bf 83}, 157 (2011). 
\bibitem{06Dil} I. Dillmann, M. Heil, F. K{\"a}ppeler, R. Plag, T. Rauscher, F-K. Thielemann, ``KADoNiS-The Karlsruhe Astrophysical Database of Nucleosynthesis in Stars," {\sc AIP Conference Proceedings} {\bf 819}, 123 (2006); Downloaded from  $\langle$http://www.kadonis.org$\rangle$ on December 4, 2017.


\bibitem{15Pri}  B.~Pritychenko, ``Calculations of Nuclear Astrophysics and Californium Fission Neutron Spectrum Averaged Cross Section Uncertainties using ENDF/B-VII.1, JEFF-3.1.2, JENDL-4.0 and Low-Fidelity Covariances," {\sc  Nucl. Data Sheets} {\bf 123}, 119 (2015).
\bibitem{06Man} W.~Mannhart, ``Response of activation reactions in the neutron field of californium-252 spontaneous fission," Report STI/DOC/10-452, 30 (2006).
\bibitem{94Tru} A.M.~Trufanov, G.N.~Lovchikova, G.N.~Smirenko, A.V.~Polyakov, V.A.~Vinogradov, ``Spectrum of $^{235}$U(n,f) Neutrons for 5 MeV Primary Neutrons," {\sc At. Energy} {\bf 76}, 195 (1994).
\bibitem{85Gra} A.F.~Grashin, M.V.~Lepeshkin, ``New Formula for the Spectrum of Prompt Neutrons from Fission," {\sc At. Energiya} {\bf 58}, 59 (1985). 
\bibitem{98Mad} D.G.~Madland, ``Theory of Neutron Emission in Fission," Los Alamos National Laboratory Report LA-UR-98-797, July 1998.
\bibitem{10Ta} P.~Talou, T.~Kawano, D.G.~Madland, A.C.~Kahler, ``Uncertainty Quantification of Prompt Fission Neutron Spectrum for n(0.5 MeV) + $^{239}$Pu," {\sc Nucl. Sci. Eng.} {\bf 166}, 254 (2010).

\bibitem{78Wo} S.E.~Woosley, W.A.~Fowler, J.A.~Holmes, B.A.~Zimmerman, ``Semiempirical thermonuclear reaction-rate data for intermediate-mass nuclei," {\sc At. Data Nucl. Data Tables} {\bf 22}, 371 (1978).
\bibitem{00Ra} T.~Rauscher, F.-K.~Thielemann, ``Astrophysical Reaction Rates From Statistical Model Calculations,"  {\sc At. Data Nucl. Data Tables} {\bf 75}, 1 (2000).
\bibitem{08Go} S.~Goriely, S.~Hilaire, A.J.~Koning, ``Improved predictions of nuclear reaction rates with the TALYS reaction code for astrophysical applications," {\sc Astr. Astrophys.} {\bf 487}, 767 (2008).
\bibitem{20Pr} B.~Pritychenko, ``Calculations of astrophysical reaction rates using ENDF/B-VIII.0 library," {\sc Nucl. Data Sheets} {\bf 167}, 76 (2020).

\bibitem{98Gre} N.~Grevesse, A.J.~Sauval, `` Standard Solar Composition," {\sc Sci. Rev.} {\bf 85}, 161 (1998).
\bibitem{57Bur}  E.M.~Burbidge, G.R.~Burbidge, W.A.~Fowler, F.~Hoyle,  ``Synthesis of the Elements in Stars," {\sc Rev. Mod. Phys.} {\bf 29}, 547 (1957).
\bibitem{57Cam} A.G.W.~Cameron,  ``Stellar Evolution, Nuclear Astrophysics and Nucleogenesis,"  AECL-454, CRL-41 (1957).
\bibitem{74Cla} D.A.~Clayton, R.A.~Ward,  ``s-Process Studies: Exact Evaluation of an Exponential Distribution of Exposures," {\sc Astrophys. Journal} {\bf 193}, 397 (1974).
\bibitem{82Kap} F.~K{\"a}ppeler, H.~Beer, K.~Wisshak, D.D.~Clayton, R.L.~Macklin, R.A.~Ward,  ``s-Process Studies In the Light of New Experimental Cross Sections: Distribution of Neutron Fluences and r-Process Residuals,"  {\sc Astrophys. Journal} {\bf 257}, 821 (1982).
\bibitem{88Rol} C.E.~Rolfs and W.S. ~Rodney, {\sc Cauldrons in the Cosmos}, The University of Chicago Press (1988).
\bibitem{81Cam} A.G.W.~Cameron, ``Essays in Nuclear Astrophysics," ed. C.A. Barnes, D.D. Clayton, and D.N. Schramm, Cambridge: Cambridge University Press (1981). 


 \bibitem{09Lod} K.~Lodders, H.~Palme, H.-P.~Gail, ``Abundances of the elements in the solar system," In Landolt-B{\" o}rnstein,  New  Series,  Vol.  VI/4B,  Chap.  4.4,  J.E. Tr{\"u}mper  (ed.),  Berlin,  Heidelberg,  New  York:  Springer-Verlag, p. 560-630 (2009). 
 \bibitem{99Arl} C.~Arlandini, F.~K{\"a}ppeler, K.~Wisshak, R.~Gallino,  M.~Lugaro, M.~Busso, O.~Sraniero, ``Neutron Capture in Low-Mass Asymptotic Giant Branch Stars: Cross Sections and Abundance Signatures," {\sc Astrophys. J.} {\bf 525}, 886 (1999).
 \bibitem{99Gor} S.~Goriely, ``Uncertainties in the solar system r-abundance distribution," {\sc Astron. Astrophys.} {\bf 342}, 881 (1999). 
\bibitem{07Arn} M.~Arnould, S.~Goriely, K.~Takahashi,  ``The r-process of stellar nucleosynthesis: Astrophysics and nuclear physics achievements and mysteries," {\sc Phys. Rept.} {\bf 450}, 97 (2007).























 
















































 
\end{thebibliography}
\end{document}